\documentclass[allclo]{FBSart}
\usepackage{amsfonts}
\usepackage{amssymb}

\title{Complex scaling of the hyper-spheric coordinates and Faddeev
equations}

\author{D.V.~Fedorov$^a$, E.~Garrido$^b$, A.S.~Jensen$^a$
\\$^a$ {\small
Department of Physics and Astronomy, Aarhus University, DK-8000
Aarhus C, Denmark}
\\$^b$ {\small
Instituto de Estructura de la Materia, CSIC-Serrano 123, E-28006,
Madrid, Spain}
}
\institute{}
\runningauthor{D.V.~Fedorov, E.~Garrido, and A.S.~Jensen}
\runningtitle{Complex scaling of the hyper-spheric coordinates...}

\begin{document}
\maketitle
\begin{abstract}
We implement complex scaling of Faddeev equations using hyper-spheric
coordinates and adiabatic expansion.  Complex scaling of coordinates
allows convenient calculations of three-body resonances.  We derive
the necessary equations and investigate the adiabatic spectrum at
large distances. We illustrate the viability of the implementation by
calculations of several three-body resonances: a $0^+$ resonance in a
model benchmark system of three identical bosons; the $2^+$ resonance
in the $^6$He nucleus within the $\alpha+n+n$ model; and the two $0^+$
resonances in $^{12}$C within the three-$\alpha$ model.

\end{abstract}

\section{Introduction}

Although the quantum-mechanical three-body problem has a long history
\cite{bethe57} it is still generally considered as ``difficult".
Recently an effective symbiotic method was introduced \cite{nielsen01}
where the Faddeev equations \cite{faddeev61} are solved using the
hyper-spheric adiabatic expansion \cite{macek68}.  The method allows
analytic asymptotic solutions and thus is especially suitable to describe
weakly bound and spatially extended systems.  During the last years it
has been successfully employed to investigate ground state properties and
fragmentation reactions of three--body halo nuclei $^6$He and $^{11}$Li
\cite{fed95,gar98,gar01}.

Investigation of three-body systems in basically all fields of
physics led very soon to the question of the three--body continuum
spectrum. Investigations of three-body resonances is today a hot topic
in both experimental and theoretical physics.

In the early 70's the complex scaling method was introduced
\cite{agu71,bal71} to calculate resonances, where the radial coordinate,
and therefore the Hamiltonian, is rotated into the complex plane. This
transformation does not change the energies of the bound states,
while resonances appear as additional bound states with
complex energy values. Thus the usual numerical techniques
used to compute bound states can also be employed to
calculate resonances \cite{ho83,kruppa88,moi98,csoto00} (see also
\cite{witala,kolganova,myo,alferova} for recent applications to three-body
systems).

A popular alternative coordinate space method to calculate resonances
is the complex energy method where the poles of the $S$-matrix are
searched for. This method requires matching of the calculated wave
function with a certain large distance asymptotics, typically a Bessel
function, but sometimes a more complicated function. Furthermore the
asymptotic functions diverge exponentially, and an accurate
calculation at large distances becomes soon a delicate numerical
task, especially for the long-range Coulomb potentials \cite{fed96}. 

The great advantage of the complex scaling method is that it is
possible to avoid matching at large distances and to employ instead a
simpler boundary condition of vanishing wave function or, alternatively,
a basis of square integrable functions.

In this paper we implement the complex scaling method within the
hyper-spheric adiabatic approach to the Faddeev equations.  We first
derive the necessary equations and investigate the asymptotic adiabatic
spectrum.  We then illustrate the viability of the method by applications
to three systems: i) a model benchmark system of three bosons; ii) the
halo nucleus $^{6}$He within the three-body $\alpha+n+n$ model; and iii)
the three-$\alpha$ system where the long range Coulomb interaction is
of importance.

\section{Complex scaling of hyper-spheric
coordinates}

In this chapter we reintroduce some pertinent definitions from the
complex scaling and hyper-spheric adiabatic methods, perform the scaling
of hyper-spheric coordinates, derive the formulae necessary for the
numerical implementation of the method, and discuss the large-distance
asymptotics of the hyper-angular adiabatic spectrum.

\subsection{Complex scaling in a two-body system}

A resonance or, in the stationary scattering language, a decaying
meta-stable state corresponds to a pole $k_0=q-i\gamma$ of the $S$-matrix
located in the IV-th quadrant of the complex momentum. The corresponding
radial wave-function $f_{k_0}$ asymptotically grows exponentially
\begin{equation}
f_{k_{0}}\propto e^{iqr}e^{\gamma r}.  \label{egr}
\end{equation}
The complex scaling method is designed to avoid this exponential growth by
scaling the coordinate $r$ with a complex exponent $e^{i\theta }$
\begin{equation}
r\rightarrow re^{i\theta }.  \label{ret}
\end{equation}
This changes the asymptotics (\ref{egr}) of the radial wave-function into 
\begin{equation}  \label{k0rs}
f_{k_{0}}\propto
e^{i|k_{0}|r\cos (\theta -\varphi )}
e^{-|k_{0}|r\sin (\theta -\varphi )},
\end{equation}
where $-\varphi$ is the argument of the complex momentum $k_{0}$ ($
k_{0}=|k_{0}|e^{-i\varphi }$). If $\theta >\varphi $ this wave-function is
exponentially vanishing and the resonance effectively becomes a bound
state and thus numerical methods designed for bound state problems could
also be employed for resonances.

The exponentially vanishing wave-function (\ref{k0rs}) allows a simple
bounding box boundary condition 
\begin{equation}
f(r_{\max })=0,  \label{frmax}
\end{equation}
provided $r_{\max }$ is sufficiently large. The continuum spectrum solution
for the complex scaled radial function at large distances is 
\begin{equation}
f(r)\propto \sin (kre^{i\theta }+\delta ),
\end{equation}
where $\delta$ is the phase shift. The boundary condition (\ref{frmax})
with this function then leads to a trade-mark rotated spectrum of the
continuum states lying along a line rotated by $-\theta $ ($-2\theta $
in the energy plane)
\begin{equation}
k_{n}\simeq \frac{\pi n}{r_{\max }}e^{-i\theta },\;n=1,2,... \label{kn2}
\end{equation}

Although the wave-function (\ref{k0rs}) is exponentially vanishing,
the fall-off rate is controlled by the parameter $\sin (\theta -\varphi )$
and therefore $r_{\max }$ must satisfy
\begin{equation}
|k_{0}|r_{\max }\sin (\theta -\varphi )\gg 1, 
\end{equation}
and thus is typically larger than for bound states, where only
\begin{equation}
|k_{0}|r_{\max }\gg 1 
\end{equation}
is needed.  Also for exponential, $\exp(-r/b)$, and gaussian,
$\exp(-r^2/b^2)$, potentials the range $b$ of the potential is
increased after complex scaling as, correspondingly, $b/\cos\theta$
and $b/\sqrt{\cos 2\theta}$.  For gaussian potentials the scaling angle
therefore can not exceed $\theta=\pi/4$.

This increase of the range of the scaled potential is counterproductive
as it necessitates numerical calculations to larger distances.

\subsection{Hyper-spheric adiabatic approach}

We first define the usual Jacobi coordinates $\mathbf{x}_{i}$ and $\mathbf{y}
_{i}$ in terms of masses $m_{i}$ and coordinates $\mathbf{r}_{i}$ of the
three particles ($i$=1,2,3) 
\begin{eqnarray}
\mathbf{x}_{i}&=&\sqrt{\mu _{i}}(\mathbf{r}_{j}-\mathbf{r}_{k})\;,\; \\
\mathbf{y}_{i}&=&\sqrt{\mu _{(jk)i}}\left( \mathbf{r}_{i}-\frac{m_{j}\mathbf{r}
_{j}+m_{k}\mathbf{r}_{k}}{m_{j}+m_{k}}\right) \;, \label{jac}\\
\mu _{i}&=&\frac{1}{m}\frac{m_{j}m_{k}}{m_{j}+m_{k}}\;,\;\mu _{(jk)i}=\frac{
1}{m}\frac{m_{i}(m_{j}+m_{k})}{m_{i}+m_{j}+m_{k}} \nonumber
\end{eqnarray}
where $\{i,j,k\}$ is a cyclic permutation of \{1,2,3\} and $m$ is an
arbitrary mass. The hyper-radius $\rho $ and the hyper-angles $\alpha _{i}$
are now defined as
\begin{equation}\label{rho}
\rho \sin (\alpha _{i})=x_{i}\;,\;\;\rho \cos (\alpha _{i})=y_{i}\;. 
\nonumber
\end{equation}
A set of hyper-spheric coordinates consists of the hyper-radius $\rho $ and
a set of five angles $\Omega _{i}$ -- the hyper-angle $\alpha _{i}$ and the
four directional angles of $\mathbf{x}_{i}$ and $\mathbf{y}_{i}$. Here
$i$ distinguishes different sets of Jacobi coordinates.

The kinetic energy operator $T$ is then given as 
\begin{eqnarray}
T=T_{\rho }+\frac{\hbar ^{2}}{2m\rho ^{2}}\Lambda^2 \;,\; \\
T_{\rho }=-\frac{
\hbar ^{2}}{2m}\left( \rho ^{-5/2}-\frac{\partial ^{2}}{\partial \rho ^{2}}
\rho ^{5/2}-\frac{1}{\rho ^{2}}\frac{15}{4}\right) \;,  \label{def} \\
\Lambda^2 =-\frac{1}{\sin (2\alpha _{i})}\frac{\partial ^{2}}{\partial
\alpha _{i}^{2}}\sin (2\alpha _{i})-4
+\frac{l_{x_{i}}^{2}}{\sin ^{2}(\alpha
_{i})}+\frac{l_{y_{i}}^{2}}{\cos ^{2}(\alpha _{i})}\;,
\end{eqnarray}
where $\mathbf{l}_{x_{i}}$ and $\mathbf{l}_{y_{i}}$ are the angular momentum
operators related to $\mathbf{x}_{i}$ and $\mathbf{y}_{i}$.

The hyper-spheric adiabatic method allows effective solution of the
three-body problem with the Hamiltonian 
\begin{equation}
H=T_{\rho }+\frac{\hbar ^{2}}{2m\rho ^{2}}\Lambda^2 +\sum_{i=1}^{3}V_{i}, 
\end{equation}
where $V_{i}$ is the two-body potential between particles $j$ and $k$. The
coordinates are divided into the ''slow'' hyper-radius $\rho $ and ''fast''
hyper-angles $\Omega $ (here we refer to any of the three sets of angles).
For every fixed hyper-radius the eigenvalue problem
\begin{equation}\label{lambda}
\left[ \Lambda^2 +\frac{2m\rho ^{2}}{\hbar ^{2}}\sum_{i=1}^{3}V_{i}\right]
\Phi _{n}=\lambda _{n}\Phi _{n},
\end{equation}
is solved and the eigenvalues $\lambda _{n}(\rho )$ and
eigen-functions $\Phi _{n}(\rho ,\Omega )$ are calculated as function
of $\rho$. The total wave-function $\Psi $ is then expanded in terms of
these eigen-functions
\begin{equation}\label{psitot}
\Psi =\frac{1}{\rho ^{5/2}}\sum_{n}f_{n}(\rho )\Phi _{n}(\rho ,\Omega ). 
\end{equation}
The expansion coefficients $f_{n}(\rho )$ satisfy a coupled system of
differential equations where the angular eigenvalues serve as effective
potentials 
\begin{equation}
\left[ -\frac{\partial ^{2}}{\partial \rho ^{2}}+\frac{\lambda _{n}+15/4}{
\rho ^{2}}-\frac{2mE}{\hbar ^{2}}\right] f_{n}=
\sum_{n^{\prime }}\left[
2P_{nn^{\prime }}\frac{\partial }{\partial \rho }+Q_{nn^{\prime }}\right]
f_{n^{\prime }},  \label{radial}
\end{equation}
where $E$ is the three-body energy and $P$ and $Q$ are the non-adiabatic
terms defined as
\begin{eqnarray}\label{pq}
P_{nn^{\prime }}(\rho ) &=&\int \Phi^{\dagger} _{n}(\rho ,\Omega )
\frac{\partial }{ \partial \rho }\Phi _{n\prime}(\rho ,\Omega )d\Omega , \\
Q_{nn^{\prime }}(\rho ) &=&\int \Phi^{\dagger} _{n}(\rho ,\Omega )
\frac{\partial ^{2}}{\partial \rho ^{2}}\Phi _{n\prime}(\rho ,\Omega )
d\Omega , \nonumber
\end{eqnarray}
where $\dagger$ denotes hermitian conjugation.

\subsection{Scaling of hyper-spheric coordinates}

We now perform the complex scaling of the Jacobi coordinates 
\begin{equation}
x\rightarrow xe^{i\theta },\;y\rightarrow ye^{i\theta }.
\end{equation}
The hyper-radius $\rho$ is then scaled with the same angle $\theta$
while the hyper-angles $\Omega$ remain real as they only depend upon
ratios of the Jacobi coordinates
\begin{equation}
\rho \rightarrow \rho e^{i\theta },\;\Omega \rightarrow \Omega .
\end{equation}

We shall in the following always maintain real $\rho $ writing the complex
rotated variable explicitly as $\rho e^{i\theta }$. The angular eigenvalue
problem (\ref{lambda}) acquires then complex potentials 
\begin{eqnarray}\label{clambda}
\left[ \Lambda^2 +e^{2i\theta }\frac{2m\rho ^{2}}{\hbar ^{2}}
\sum_{i=1}^{3}V_{i}(\rho e^{i\theta },\Omega )\right] \Phi _{n}
\nonumber \\ =\lambda_{n}\Phi_{n},
\end{eqnarray}
and therefore the eigenvalues $\lambda_{n}$ and eigen-functions $\Phi_{n}$
also become complex.

We solve the complex-scaled eigenvalue problem (\ref{clambda}) by using
the hyper-spheric harmonics as a basis set \cite{nielsen01}.
Since the hyper-angles are not affected by the complex scaling these
functions remain real. The operator $\cal H$ at the left hand side of the
eigenvalue equation (\ref{clambda}),
\begin{equation}\label{a}
\left[ \Lambda^2 +e^{2i\theta }\frac{2m\rho ^{2}}{\hbar ^{2}}
\sum_{i=1}^{3}V_{i}(\rho e^{i\theta },\Omega )\right] \equiv \cal H
, \end{equation}
is in this basis a symmetric complex matrix, $\cal H^{T}=\cal H$ (here $^T$
denotes transposition). A non-scaled hamiltonian, $\theta=0$, is a real
symmetric matrix, that is a hermitian matrix.

A complex matrix $\cal H$ has in general different left, $\chi$, and right,
$\Phi$, eigenvectors corresponding to a given eigenvalue $\lambda$
\begin{eqnarray}
{\cal H}\Phi &=&\lambda \Phi ,\\
\chi^{\dagger}{\cal H} &=&\lambda \chi^{\dagger} . \nonumber
\end{eqnarray}
The left $\chi_{n}$ and right $\Phi_{n^{\prime }}$ eigenvectors
corresponding to different eigenvalues $\lambda _{n}$ and $\lambda
_{n^{\prime }}$ are orthogonal in the following way
\begin{equation}
\chi_{n}^{\dagger}\Phi_{n^{\prime }}=\delta _{nn^{\prime }}.  \label{v+u}
\end{equation}
For a symmetric matrix there is an apparent connection between the left
and right eigenvalues
\begin{equation}
\chi^{\dagger}=\Phi^{T}.
\end{equation}
Therefore for the complex scaled angular eigen-vectors the scalar product
rule takes a somewhat unusual form
\begin{equation}
\Phi_{n}^{T}\Phi_{n^{\prime }}=\delta _{nn^{\prime }},  \label{utu}
\end{equation}
and thus the hermitian conjugation in the definitions of $P$ and $Q$ in
(\ref{pq}) must be substituted by transposition in the case of complex
scaling.

The hyper-radial equations (\ref{radial}) remain largely unchanged, 
but, of course, the eigenvalues $\lambda _{n}$ together with the
non-adiabatic terms $P$ and $Q$ are now complex.

Similar to the two-body case a three-body resonance corresponds to a pole
of the three-body $S$-matrix at a momentum $\kappa=q-i\gamma$
($\kappa =\sqrt{2mE/\hbar ^{2}}$). At this point all hyper-radial
functions $f_n$ have the diverging asymptotics
\begin{equation}
f_n \propto e^{iq\rho}e^{\gamma\rho}.
\end{equation}

The complex scaling of the hyper-radius leads in complete analogy to the
two-body case to an oscillating and exponentially vanishing resonance
function

\begin{equation}\label{frhosin}
f_n\propto
e^{i|k_{0}|\rho\cos (\theta -\phi )}
e^{-|k_{0}|\rho\sin (\theta -\phi )},
\end{equation}
and again the boundary condition
\begin{equation}
f_n(\rho_{max}=0)
\end{equation}
leads to the rotated discretized continuum states
\begin{equation} \label{kapn}
\kappa_j\simeq \frac{\pi j}{\rho_{\max }}e^{-i\theta },\;j=1,2,...
\; .
\end{equation}

\subsection{Faddeev equations.}

The angular wave-function $\Phi$ is now represented as a
sum of three components $\Phi=\phi^{(1)}+\phi^{(2)}+\phi ^{(3)}$ which
satisfy the three Faddeev equations
\begin{eqnarray}\label{cfad}
\Lambda^2 \phi^{(k)} +
e^{2i\theta}\frac{2m\rho^2}{\hbar^2}V_{k}(\rho e^{i\theta },\Omega)
\Phi
=\lambda\phi^{(k)},
\end{eqnarray}
where $k$=1,2,3. One can easily verify that the sum of these three Faddeev
equations is equivalent to the Schr\"{o}dinger equation (\ref{clambda}).

The Faddeev equations (\ref{cfad}) can be conveniently written in a
matrix form as
\begin{equation}\label{flam}
F\phi = \lambda \phi,
\end{equation}
where $\phi$ is a column vector of the three Faddeev components
\begin{equation}\label{fadvec}
\phi =\left( \begin{array}{l}
\phi ^{(1)} \\ 
\phi ^{(2)} \\ 
\phi ^{(3)}
\end{array} \right)
\end{equation}
and the Faddeev operator $F$ is equal
\begin{equation}
F = T + V R,
\end{equation}
where $T$ is the kinetic energy matrix 
\begin{equation}
T=\left( \begin{array}{lll}
\Lambda^2 & 0 & 0 \\ 
0 & \Lambda^2 & 0 \\ 
0 & 0 & \Lambda^2
\end{array} \right) ,
\end{equation}
$V$ is the potential matrix 
\begin{equation}
V=\left( 
\begin{array}{lll}
V_{1} & 0 & 0 \\ 
0 & V_{2} & 0 \\ 
0 & 0 & V_{3}
\end{array}
\right) .
\end{equation}
and $R$ is the projection matrix
\begin{equation}
R =\left( \begin{array}{lll}
1 & 1 & 1 \\ 
1 & 1 & 1 \\ 
1 & 1 & 1
\end{array} \right).
\end{equation}

If we again use hyper-spheric harmonics as a basis set then each of the
elements in the above matrices becomes in turn a matrix in this basis.

The matrix $R$ acting on the Faddeev vector (\ref{fadvec}) apparently
produces a vector with the total function $\Phi$ as the three
components. The matrix $R$ has the following properties
\begin{eqnarray}
R^{2} &=&3R, \\
R^{T} &=&R  \nonumber
\end{eqnarray}

The scalar product rule (\ref{utu}) for the Faddeev functions becomes 
\begin{eqnarray}
\phi _{n}^{T}R^{T}R\phi _{n^{\prime }} &=&\delta _{nn^{\prime
}}\;\Rightarrow  \\
\phi _{n}^{T}R\phi _{n^{\prime }} &=&\frac{1}{3}\delta _{nn^{\prime }}. 
\nonumber
\end{eqnarray}

The factor $1/3$ can be hidden into the normalization by
redefining the wave-functions such that
\begin{equation}
\phi _{n}^{T}R\phi _{n^{\prime }}=\delta _{nn^{\prime }}.  \label{ftf}
\end{equation}
This product rule should be used in the definitions (\ref{pq}) of the
non-adiabatic terms $P$ and $Q$ if one uses the Faddeev three-component
functions:
\begin{eqnarray}\label{pq2}
P_{nn^{\prime }}(\rho ) &=&\int \phi^{T} _{n}R
\frac{\partial \phi _{n\prime}}{ \partial \rho }d\Omega , \\
Q_{nn^{\prime }}(\rho ) &=&\int \phi^{T} _{n}R
\frac{\partial ^{2}\phi _{n\prime}}{\partial \rho ^{2}}
d\Omega . \nonumber
\end{eqnarray}

Actually, the product rule with the $R$-matrix (\ref{ftf}) represents
the ordinary product rule for the Faddeev operator $F$.  Although the
Faddeev matrix is not hermitian, it has the important symmetry property

\begin{equation}
RF=F^{T}R,  \label{rf=ftr}
\end{equation}
which leads to a simple connection between the left, $\varphi$, and right,
$\phi$, eigenvectors of the matrix F
\begin{equation}\label{vu}
\varphi^{\dagger}=\phi^{T}R, 
\end{equation}
which immediately leads to (\ref{ftf}).

Equation (\ref{vu}) can be proved by multiplying (\ref{flam}) with $R$
from the left, transposing it and using the relation (\ref{rf=ftr}).
Apparently, since we did not assume that the matrix $V$ is real, the
complex scaling does not affect the validity of the above expressions.

Note that for the non-scaled real $F$ the property (\ref{rf=ftr})
means that the eigenvalues of $F$ are real although the matrix itself
is not hermitian.

\subsection{Phase of the angular eigen-functions}

The complex eigen-functions $\Phi _{n}$ obtained from the complex scaled
eigenvalue equation (\ref{clambda}) are as usual defined up to a phase
factor.  We can therefore multiply the eigen-function $\Phi _{n}$ by
an arbitrary phase function $e^{i\chi_n(\rho)}$, where $\chi_n$ is real.
This changes the non-adiabatic terms $P$ and $Q$ in (\ref{pq}) as
\begin{eqnarray}
P_{nn^{\prime }} &\rightarrow &P_{nn^{\prime }}+i\chi _{n^{\prime }}^{\prime
}\delta _{nn^{\prime }}, \label{pp} \\
Q_{nn^{\prime }} &\rightarrow &Q_{nn^{\prime }}+2i\chi _{n^{\prime
}}^{\prime }P_{nn^{\prime }}
+\left( i\chi _{n^{\prime }}^{\prime \prime
}-\chi _{n^{\prime }}^{\prime 2}\right) \delta _{nn^{\prime }},  \label{qq}
\end{eqnarray}
(where primes denote differentiation with respect to $\rho$) and
correspondingly, via equation (\ref{radial}), leads to different radial
functions $f_n$. However, the total wave-function $\Psi\propto\Phi_n f_n$
in (\ref{psitot}) does not change as the new set of radial functions
merely corresponds to a trivial phase shift

\begin{equation} f_{n}\rightarrow
e^{-i\chi _{n}(\rho)}f_{n} \; ,
\end{equation}
such that the product of the angular and the radial function remains
unchanged
\begin{equation}
\Phi _{n}f_{n}\rightarrow e^{i\chi_n}\Phi_ne^{-i\chi _{n}(\rho)}f_{n}
=\Phi _{n}f_{n}. 
\end{equation}

Indeed the right hand side of the radial equations (\ref{radial}) transforms
as 
\begin{eqnarray}
\left[ 2P_{nn^{\prime }}\frac{\partial }{\partial \rho }+Q_{nn^{\prime
}}\right] f_{n^{\prime }} \rightarrow 
\nonumber\\
e^{-i\chi _{n^{\prime }}}\big(\left[ 2P_{nn^{\prime }}\frac{\partial }{
\partial \rho }+Q_{nn^{\prime }}\right] f_{n^{\prime }}
+\left( \chi
_{n^{\prime }}^{\prime 2}+i\chi _{n^{\prime }}^{\prime \prime }\right)
\delta _{nn^{\prime }}f_{n^{\prime }}+2i\chi _{n^{\prime }}^{\prime }\delta
_{nn^{\prime}}\frac{\partial}{\partial\rho}f_{n^{\prime }}\big)
\end{eqnarray}
The last two terms, however, precisely cancel with the corresponding terms
from the second derivative at the left hand side of the radial equations
(\ref{radial})
\begin{equation}
\frac{\partial ^{2}}{\partial \rho ^{2}}f_{n}\rightarrow \frac{\partial ^{2}
}{\partial \rho ^{2}}e^{-i\chi _{n}}f_{n}=
e^{-i\chi _{n}}\big( \frac{\partial ^{2}}{\partial \rho ^{2}}f_{n}
-i\chi _{n}^{\prime \prime
}f_{n}-\chi _{n}^{\prime 2}f_{n}-2i\chi _{n}^{\prime }
\frac{\partial}{\partial\rho}f_{n}\big) . 
\end{equation}

Thus the phase of the hyper-angular eigen-functions can be chosen
arbitrarily as the total wave-function is independent of this phase.

In the non-scaled case the diagonal elements $P_{nn}$ are identically
equal zero. It is convenient to chose the phase such that this property
holds also for the scaled case. One can easily verify from (\ref{pp})
that the phase
\begin{equation}
\chi _{n}(\rho )=\exp \left( i\int_{0}^{\rho }P_{nn}(\rho ^{\prime })d\rho
^{\prime }\right) , 
\end{equation}
precisely leads to a vanishing diagonal $P$ as in the non-complex-scaled
case.

The freedom in choosing the phase can also be exploited in order to
eliminate a given non-diagonal term $Q_{nn^{\prime }}.$ The phase should in
this case instead be chosen as 
\begin{equation}
\chi _{n^{\prime }}^{(n)}(\rho )=\exp \left( \frac{i}{2}\int_{0}^{\rho
}P_{nn^{\prime }}^{-1}(\rho ^{\prime })Q_{nn^{\prime }}(\rho ^{\prime
})d\rho ^{\prime }\right) ,
\end{equation}
as seen from (\ref{qq}).

\subsection{Asymptotics of the eigenvalues}
In the angular eigenvalue problem (\ref{clambda}) the two-body potentials
are multiplied by $\rho^2$. At the origin, $\rho$=0, they therefore
vanish from the equations and the eigenvalues then start from the free
spectrum $K(K+4)$, where $K$ is even or odd integer. The next term in
the $\rho$ expansion is proportional to $\rho^2$ as in the non-scaled
case \cite{nielsen01}.

The large distance asymptotics can be investigated by use of zero-range
potentials.  For our purpose it is also sufficient to assume that only
one pair of particles interact with each other.  This simple model
will show all different types of asymptotics except for the Efimov
effect \cite{fedorov93}.  However the latter is not the subject of the
current investigation.

In this case the physical (non-spurious \cite{nielsen01}) solution to
the Faddeev equations (\ref{cfad}) contains only one non-zero Faddeev
component in (\ref{fadvec}), namely the one where the two-body potential
acts.

The zero-range potential is vanishing everywhere except for the origin.
The physical (one-component) solution to the equation (\ref{cfad})
is then a simple function
\begin{equation}
\phi = \sin(\nu(\alpha-\frac{\pi}{2})),
\end{equation}
where $\nu=\sqrt{\lambda +4}$.  The zero range potential enters as a
boundary condition at the origin \cite{renorm}
\begin{equation}\label{zero}
-\nu \frac{
\cos(\nu\frac{\pi}{2})
}{
\sin(\nu\frac{\pi}{2})
}=\frac{1}{\sqrt{\mu }}\frac{\rho}{a} e^{i\theta},
\end{equation}
where $\mu$ is the reduced mass of two particles in units of $m$, and
$a$ is the two-body scattering length \footnote{The sign convention
for the scattering length is $k\cot(\delta)=1/a+O(k^{2})$}.

For $\rho\gg |a|$ this equation has several types of solutions.  First there
is a series of asymptotic solutions where $\sin(\nu\frac{\pi}{2})$ is zero
\begin{equation}\label{free}
\nu_n = 2n \;, \; \lambda_n = (2n)^2-4 \;, \; n=1,2,...
\end{equation}
which is the free hyper-spheric spectrum $K(K+4)$ with $K=2n-2$.

Another type of solution may exist when $\cot(\nu\frac{\pi}{2})$ remains
finite but the complex quantity $\nu$ instead is proportional to $\rho$.
Suppose that the scattering length is negative, $a<0$, corresponding
to a bound two-body state with binding energy $B=\hbar^2/(2m\mu a^2)$.
In this case there exists an asymptotic solution to (\ref{zero})

\begin{equation}
\nu=\frac{1}{\sqrt{\mu}}\frac{\rho}{a}e^{i(\theta+\pi/2)},
\end{equation}
since in this case $\cot(\nu{\pi\over 2})\rightarrow i$.

Therefore a bound two-body state always results in an asymptotically
quadratically diverging eigenvalue $\lambda=\nu^2-4$
\begin{equation}\label{cbound}
\lambda = -\frac{2mB}{\hbar^2}\rho ^{2}e^{2i\theta } \;.
\end{equation}

For positive scattering length, $a>0$ and $\theta<\pi/2$ only the free
asymptotic solutions (\ref{free}) to the equation equation (\ref{zero})
exist.  Positive scattering length corresponds to a resonance (more
precisely a virtual state) which is located at the complex momentum
$k=-i/a$, that is with the argument $-\pi/2$ in the complex momentum
plane. The energy of the virtual state is given by $E_0=-\hbar^2/(2m\mu
a^2)$.

When $a>0$ and the scaling angle $\theta$ is larger than $\pi/2$ yet
additional asymptotic solution to (\ref{zero}) appears

\begin{equation}
\nu=\frac{1}{\sqrt{\mu}}\frac{\rho}{a} e^{i(\theta-\pi/2)}.
\end{equation}

That is when the scaling angle $\theta$ is larger than the angle of a
resonance ($\pi/2$ in the case of a virtual state) the latter effectively
becomes a bound state resulting in an angular eigenvalue similar to
(\ref{cbound}) but with the resonance (or virtual state) energy $E_{0}$
instead of the bound state energy

\begin{equation}\label{cres}
\lambda = \frac{2mE_0}{\hbar^2}\rho^{2}e^{2i\theta} .
\end{equation}

Thus the asymptotic spectrum of hyper-angular eigenvalues primarily
consists of the free spectrum (\ref{free}). In addition each two-body
bound state gives a quadratically diverging eigenvalue (\ref{cbound}),
and finally each two-body resonance (or virtual state) whose argument is
less the scaling angle $\theta$ also provides a quadratically divergent
eigenvalue (\ref{cres}).

The eigenvalues enter the hyper-radial equations (\ref{radial}) as
effective potentials $\lambda/\rho^2$.  After divided by $\rho^2$ the
free eigenvalues (\ref{free}) converge to the three-body threshold at
zero energy.  These eigenvalues asymptotically describe the genuine
three-body states where all three particles are away from each other.
The eigenvalues (\ref{cbound}) corresponding to the two-body bound state
after  division by $\rho^2$ converge to a lower threshold at the energy
of the two-body bound state.  These eigenvalues asymptotically describe a
configuration where two of the particles are in the two-body bound state
while the third particle is in the continuum.  Finally, the two-body
resonances and virtual states which after complex scaling appear as
bound states produce the same type of asymptotic solutions as the true
bound states.
\section{Numerical examples}
In this chapter we investigate several resonances in different three-body
systems using the formulated complex-scaled hyper-spheric method
and compare with other methods. The first application is a benchmark
calculation of a resonance in a model three-boson system \cite{kolganova}
with short-range potentials. The second is a narrow 2$^+$ resonance in the
halo nucleus $^6$He. The third application is the two lowest 0$^+$
resonances in a system of three $\alpha$-particles where the long-range
Coulomb forces are particularly important.

First we solve the angular eigenvalue problem (\ref{clambda}) by
expansion in hyper-spheric harmonics and obtain thereby several
of the lowest eigenvalues $\lambda_n(\rho)$ as functions of $\rho$.
For the system of radial equations (\ref{radial}) we employ the finite
difference method. A mesh of about 100 points in hyperradius from
zero to a certain maximal distance $\rho_{max}$ is chosen such that
the density of points is highest in the regions were the non-adiabatic
terms $P$ and $Q$ contribute most. We then approximate the second and
first derivatives in (\ref{radial}) by a five point finite difference
approximation assuming that the radial functions are zero at the origin
and at $\rho_{max}$. The system of differential equations (\ref{radial})
then turns into a complex-matrix diagonalization problem. Numerical
diagonalization of the matrix gives the complex eigenvalues which include
all bound states, resonances and also discretized continuum states.
The resulting eigenvectors give the values of radial functions at the
mesh points.

The shooting method with direct Runge-Kutta integration of the radial
equations gives a better precision but requires a separate search for
each of the complex eigenvalues. Thus the preferred strategy could be
first to get an overal picture of the complex spectrum by the finite
difference method and then to calculate selected eigenvlaues with higher
precision by the shooting method.

For each of the following examples we show the angular eigenvalues, the
complex energy spectrum and the hyper-radial functions of the resonances.
As these calculation only serve the purpose of illustration we have
not optimized the hyper-radial mesh to achieve the maximum accuracy.
With about 100 hyper-radial points the accuracy is about 10~keV in the
complex energy.

\subsection{A model three-body system}

\begin{figure}
\input{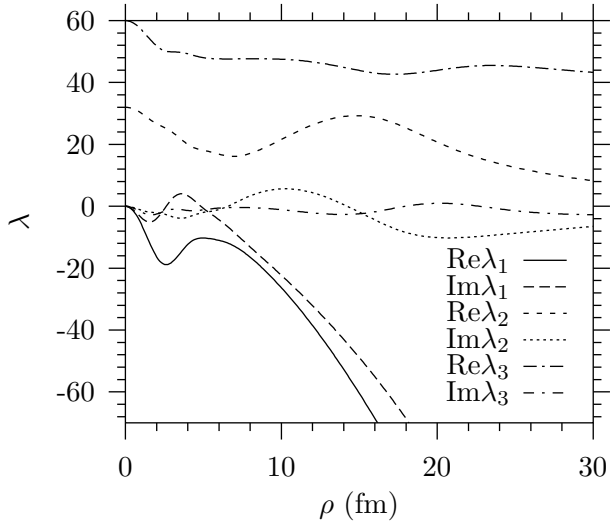}
\caption{The three lowest complex angular eigenvalues for the model
system of three identical bosons with scaling angle $\theta$=0.35. }
\label{fig-m-l}
\end{figure}

In reference \cite{kolganova} a resonance has been calculated in a model
three-body system of identical bosons with the nucleon
mass. The model potential acts only on the s-waves and consists of a
gaussian attractive pocket and a barrier
\begin{eqnarray}
V(r)=-55{\rm MeV}\exp (-0.2{\rm fm}^{-2}r^{2})\nonumber
+1.5{\rm MeV}\exp (-0.01{\rm fm}^{-2}(r-5{\rm fm})^{2}).
\end{eqnarray}
A three-body resonance was found in \cite{kolganova} at
$E=-5.9525-0.4034i$~MeV by solving two-dimensional equations.

In our calculation we adopt the nucleon mass $m$=939~MeV as the mass
unit and use $\hbar c$=197.3~MeV~fm. The three lowest angular eigenvalues
obtained by solving equation (\ref{clambda}) for the complex scaling
angle $\theta$=0.4 are shown on Fig.~\ref{fig-m-l}. At the origin the three
lowest eigenvalues are real and equal to 0, 32 and 60 which correspond to a
free hyper-spheric real spectrum of $K(K+4)$ with $K=0,4,6$. The number
$K$=2 is forbidden by the bosonic symmetry requirement.

At large distances the lowest eigenvalue diverges quadratically according
to (\ref{cbound}) due to a bound two-body state.  The other eigenvalues
return back to the free real spectrum at large distances with one of
them taking the place of the diverged lowest eigenvalue.

\begin{figure}
\input{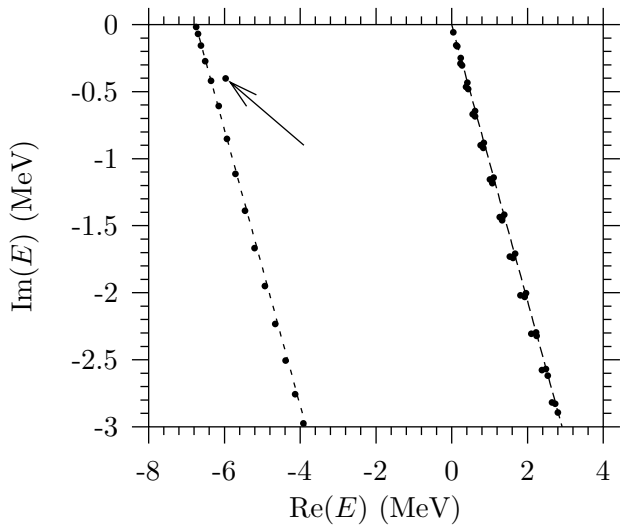}
\caption{The complex energy spectrum of the model three-boson system
with scaling angle $\theta$=0.4. The bound state
at $E=-37.22$~MeV is not shown. The resonance at $E=-5.96-0.40i$~MeV
is indicated by an arrow. The two lines, $\Im(E)=-\tan(2\theta)\Re(E)$
and $\Im(E)=-\tan(2\theta)(\Re(E)-E^{(2)})$ (where $E^{(2)}=-6.76$~MeV is the
two-body bound state energy), represent the exact rotated continuum spectrum.
}
\label{fig-m-s}
\end{figure}

The energy spectrum for this model system is shown on
Fig.~\ref{fig-m-s}. The points which form two lines represent the
discretized continuum spectrum. The line which starts from $E$=0
corresponds to the states with genuine three-body asymptotics where
all three particles are at the continuum. Another line which starts
at the two-body bound state energy $E^{(2)}=-6.76$~MeV corresponds to the
continuum states where two of the particles are confined in the bound
two-body state while the third one is in the continuum.

Using four angular eigenvalues and $\rho_{max}\sim$100~fm we find
a bound three-body ground state at $E_0=-37.22$~MeV and a resonance
at $E_1=-5.96-0.40i$~MeV in agreement with \cite{kolganova}. The
independece of the calculated energies upon the scaling angle is
illustrated in Table~\ref{tab-m}.

\begin{table}
\caption{The ground state energy $E_0$ and the resonance energy $E_1$
for different scaling angles $\theta$ for the model system}

\begin{tabular}[]{c|c|c|}
$\theta$ & $E_0$, MeV & $E_1$, MeV \\
\hline
0.30   & $-37.221$ &  $-5.968-0.400i$ \\
0.35   & $-37.220$ &  $-5.962-0.404i$ \\
0.40   & $-37.221$ &  $-5.963-0.401i$
\label{tab-m}\end{tabular}

\end{table}

\begin{figure}
\input{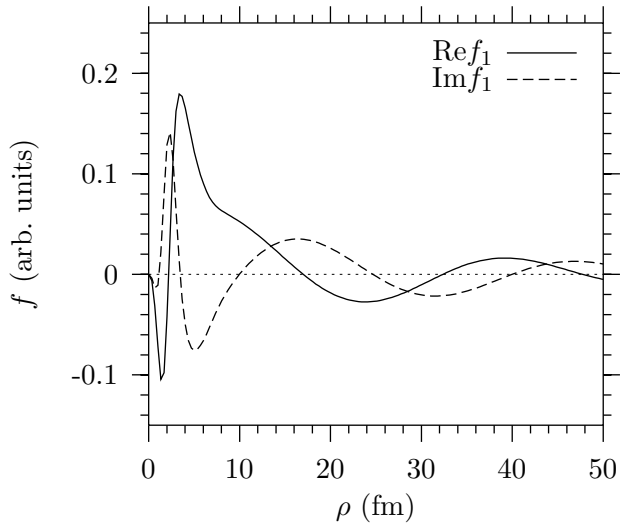}
\caption{The dominating hyper-radial function $f_1$ of the lowest
resonance in the model three-boson system. The scaling angle is
$\theta$=0.4.}
\label{fig-m-r}
\end{figure}

The radial function for the resonance is shown on Fig.~\ref{fig-m-r}.
At shorter distances the function looks like a bound first excited state
as indicated by the single node in the real part at shorter distances. At
larger distances though, unlike the bound state functions, the resonance
function oscillates with decreasing amplitude due to the complex scaling
in accordance with (\ref{k0rs}).

\subsection{Application to $^{6}$He}

In this section we apply the complex scaled adiabatic hyper-spheric
method to the 2$^+$ resonance in the halo nucleus $^6$He within the
three-body $\alpha$+n+n model.

We use the same interactions as in \cite{cobis}, namely gaussian
potentials fitted to the low energy scattering data and a phenomenological
gaussian three-body force. This force is introduced to account for
the additional polarisation of the alpha particle due to the presence
of another neutron. It simulates the small additional attraction in
$\alpha+n+n$ coming from excitations of the $\alpha$-particle.  The mass
of the $\alpha$-particle is assumed to be 4$\times$939~MeV.

\begin{figure}
\input{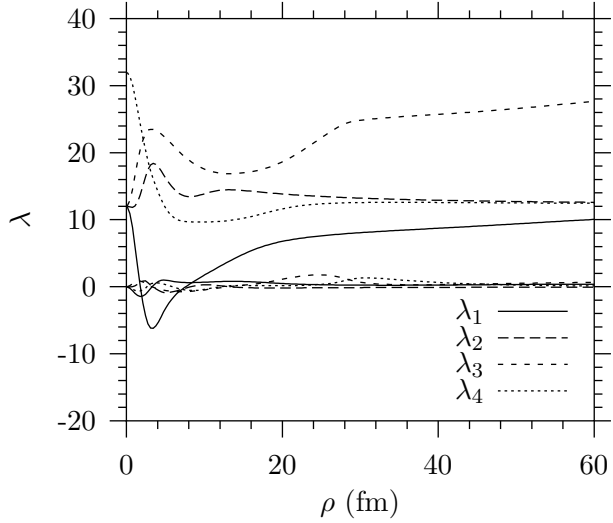}
\caption{The four lowest complex eigenvalues for the
$\alpha$+n+n system, $J^\pi$=2$^+$.  The scaling angle is $\theta$=0.2.}
\label{fig-he-l}
\end{figure}

The four lowest angular eigenvalues are shown on Fig.~\ref{fig-he-l}
for the scaling angle $\theta$=0.2.
The eigenvalues start at $\rho$=0 from the real numbers 12 and 32 which
correspond to the free real hyper-spheric spectrum of $K(K+4)$ with the
hyper-angular momentum K=2 and 4.  One of the eigenvalues then forms
an attractive pocket where the resonance resides. This is largely due
to n-$\alpha$ $p$-waves where the potential is attractive. The other
two eigenvalues relate to n-$\alpha$ $s$-waves with strong repulsion
which leads to the two repulsive eigenvalues contributing much less to
the resonance wave-function. At large distances the eigenvalues return
back to the free real spectrum as there are no two-body bound states in
this system and the scaling angle is too small for the two-body
resonances to become bound.

\begin{figure}
\input{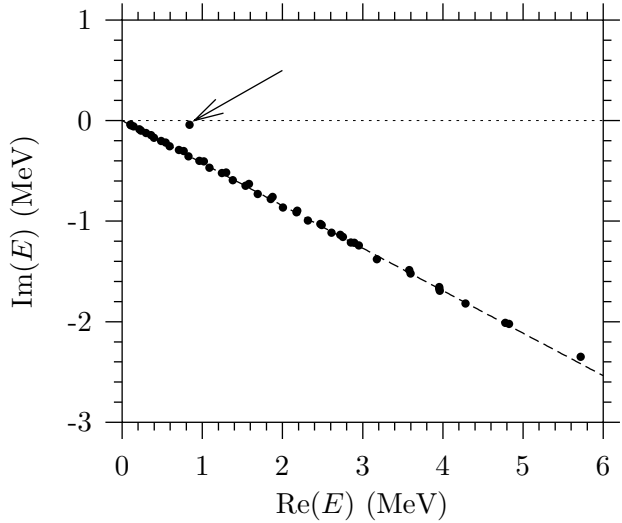}
\caption{ The complex energy spectrum of $^6$He 2$^+$ states for the
scaling angle $\theta$=0.2.  The resonance at
$E=0.842-0.041i$~MeV is indicated by an arrow. The line
$\Im(E)=-\tan(2\theta)\Re(E)$ represents the exact rotated
continuum spectrum.}
\label{fig-he-s}
\end{figure}

\begin{table}
\caption{Resonance energy $E$ for different scaling angles $\theta$ for
the $\alpha+n+n$ system}
\begin{tabular}[]{c|c}
$\theta$ & $E$, MeV \\
\hline
0.15 & $0.8419 - 0.0405i$ \\
0.20 & $0.8417 - 0.0404i$ \\
0.25 & $0.8416 - 0.0405i$
\end{tabular}
\label{tab-he}
\end{table}

The complex scaled spectrum for 2$^+$ states in $^6$He is shown on
Fig.~\ref{fig-he-s}. No bound three-body states but one narrow resonance
are present with these quantum numbers.  With four adiabatic channels and
$\rho_{max}\sim$100~fm we obtain the 2$^+$ resonance at $E=0.842-0.041i$~MeV
which is reasonably close to the result $E=0.845-0.046i$~MeV obtained in
\cite{cobis} with the complex energy method and $\rho_{max}$=180~fm. A
similar result, $E=0.81-0.065i$, has been obtained in \cite{myo},
also with complex scaling method but with a different interaction model.
The independence of the resonance energy on the scaling angle is
illustrated in Table~\ref{tab-he}.

\begin{figure}
\input{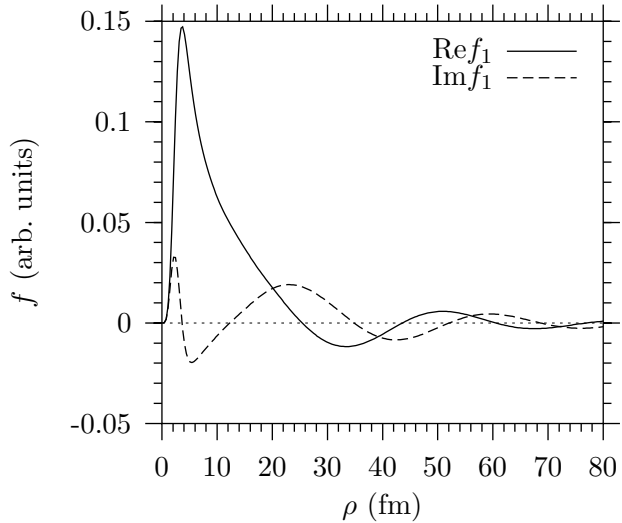}
\caption{The dominating hyper-radial function $f_1$ of the $\alpha$+n+n
system for the $J^\pi=2^+$ resonance for the scaling angle $\theta$=0.2.}
\label{fig-he-r}
\end{figure}

The largest hyper-radial function of the resonance is shown on
Fig.~\ref{fig-he-r}.  In the inner region the hyper-radial function
resembles a bound ground state function with no nodes and a considerable
concentration in the pocket region. Again we see a fingerprint of a
complex scaled resonance, namely the wave-function oscillating at large
distances with ever decreasing amplitude according to (\ref{frhosin}).

\subsection{Application to the three-$\alpha$ system}

The narrow 0$^+$ resonance in $^{12}$C nucleus with $E=0.38$~MeV and
$\Gamma=8.3\pm$1.0~eV was described quantitatively in \cite{fed96} within
a three-$\alpha$ model using the complex energy method. Here we repeat the
calculation using the same parameters but instead with the complex scaling
method. Unlike the first two numerical examples this one now includes a
long range Coulomb potential which significantly alters the asymptotics
of the wave-function \cite{fed96} compared to short range potentials.

The $\alpha$-$\alpha$ potential is taken as in \cite{fed96}
\begin{eqnarray}
V(r)&=&(125{\rm MeV}\hat{P}_{l=0}+20{\rm MeV}\hat{P}_{l=2})
e^{-r^2/(1.53{\rm fm})^2} \nonumber \\
&-&30.18{\rm MeV}e^{-r^2/(2.85{\rm fm})^2},
\end{eqnarray}
where $\hat{P}_{l}$ is the projection operator onto a state with relative
orbital momentum~$l$. The three-body force is $V_3(\rho)=-96.8{\rm
MeV}\exp(-\rho^2/(3.9{\rm fm})^2)$ where the nucleon mass $m$=939~MeV
was used as a scale in the definition (\ref{jac}-\ref{rho}) of the
hyper-radius $\rho$.

\begin{figure}
\input{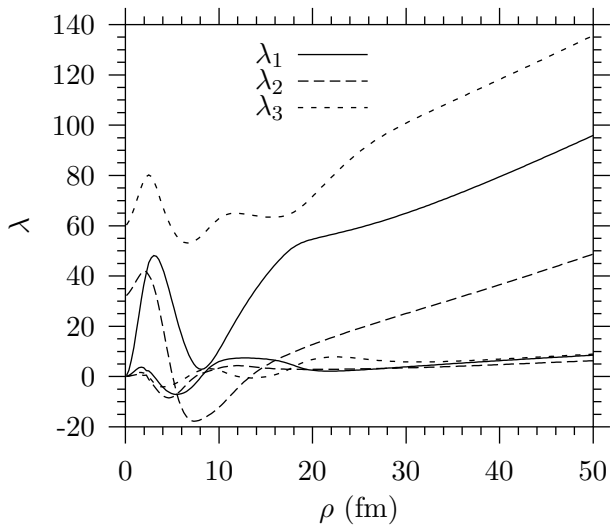}
\caption{The three lowest complex scaled angular eigenvalues
for the three-$\alpha$ system.  The scaling angle $\theta$=0.11.}
\label{fig-3a-l}
\end{figure}

The angular eigenvalues for this system are shown on
Fig.~\ref{fig-3a-l}. The lowest eigenvalue starts from zero corresponding
to $K=0$.  It contains mostly repulsive s-waves and is always positive
without any attractive pocket.  The next eigenvalue starts from
$\lambda$=32 ($K=4$) as the hyper-angular quantum number $K=2$ is forbidden
by symmetry requirements. It contains mostly attractive p-waves and does
form a relatively deep attractive pocket.

At large $\rho$ the real and imaginary parts of the eigenvalues grow
linearly instead of converging to a real constant as is the case with
short range potentials. This linear growth is due to the long range
Coulomb potential \cite{fed96} which reappears in the effective potentials
after the eigenvalue is divided by $\rho^2$.

\begin{figure}
\input{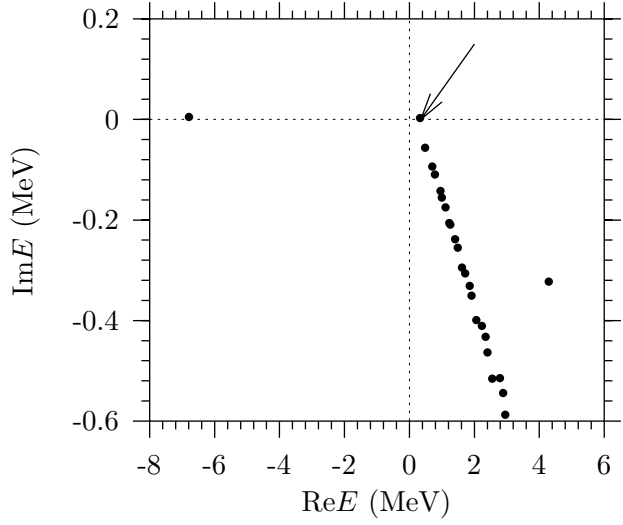}
\caption{
The complex energy spectrum of the three-$\alpha$ 0$^+$ states for the
scaling angle $\theta$=0.11.  The resonance at $E=0.33$~MeV is indicated
by an arrow.
}
\label{fig-3a-s}
\end{figure}

The energy spectrum is shown on Fig.~\ref{fig-3a-s} for the scaling
angle $\theta$=0.1 and $\rho_{max}$=60~fm. There is a bound ground
state located at $E_0=-6.79$~MeV and two resonances, the very narrow one
at $E_1=0.330$~MeV and a broader resonance at $E_2=4.30-0.32i$~MeV.

\begin{figure}
\input{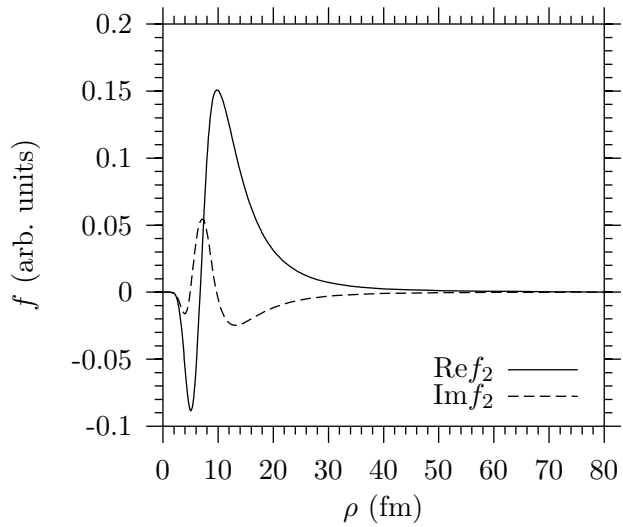}
\caption{The dominating hyper-radial function $f_2(\rho)$ for the
three-$\alpha$ system for the lowest 0$^+$ resonance and $\theta$=0.11.}
\label{fig-3a-r}
\end{figure}

It is somewhat difficult to tell the first resonance from a discretized
continuum state judging from the energy spectrum plot alone. However the
hyper-radial wave-function shown on Fig.~\ref{fig-3a-r} leaves no doubts
that this is indeed a resonance. Shown is the dominating function $f_2$
which corresponds to the second eigenvalue on Fig.~\ref{fig-3a-l} where
the attractive pocket is present. The function is again concentrated
at smaller distances and has one node in the real part corresponding
to the first excited state.  At larger distances it basically falls
off without oscillations like as if it were a true bound state. This
behavior results from the large effective Coulomb barrier provided by
the linearly growing angular eigenvalues as shown on Fig.~\ref{fig-3a-l}.

Another criterion to distinguish the resonances from the continuum states
is that when the scaling angle $\theta$ is changed the whole line of
continuum states rotates according to (\ref{kapn}) while the resonances
(and bound states) are independent of the scaling angle as illustrated in
Table~\ref{tab-3a}.

\begin{table}
\caption{The bound state energy $E_0$ and resonance energies $E_1$
and $E_2$ of the three lowest 0$^+$ states in 3$\alpha$ system}

\begin{tabular}{c|c|c|c}
$\theta$ & $E_0$, MeV & $E_1$, MeV & $E_2$, MeV \\
\hline
0.11   & $-6.789$ & 0.3300     & $4.295-0.323i$ \\
0.12   & $-6.793$ & 0.3283     & $4.294-0.323i$ \\
0.13   & $-6.790$ & 0.3297     & $4.296-0.325i$ \\
0.14   & $-6.790$ & 0.3296     & $4.300-0.321i$
\end{tabular}
\label{tab-3a}\end{table}

The energies obtained in this calculation compare well to the values of
-6.81~MeV for the ground state and 380-0.02$i$~keV for the first resonance
obtained in \cite{fed96} with the complex-energy method.  The source of
the difference is that in \cite{fed96} only two of the lowest angular
eigenvalues were used while here we use as many eigenvalues as needed for
convergence -- typically 4-5. Again for these numerical illustrations
we have more or less arbitrarily chosen about 100 mesh points in the
hyperradius for all examples. No fine tuning has been made in order to
achieve higher accuracy. The present numerical error is estimated to
be about 10~keV and, therefore, the width of the order of 0.01~keV of
the first extremely narrow resonance is beyond the capabilities of an
untuned mesh.

\begin{figure}
\input{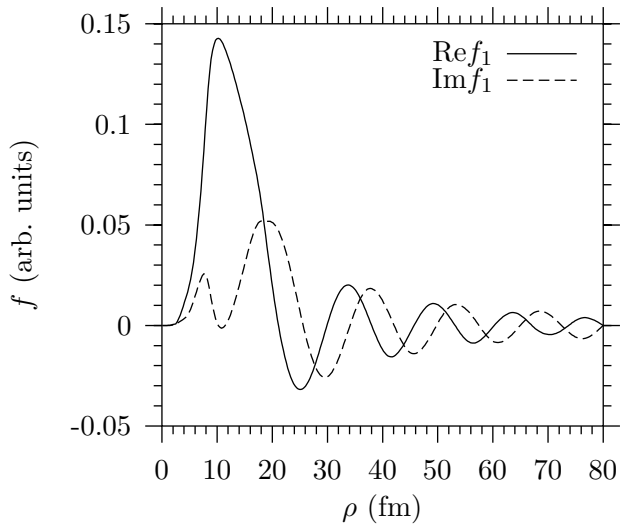}
\caption{The dominating hyper-radial function $f_1(\rho)$ for the
three-$\alpha$ system for the second 0$^+$ resonance and $\theta$=0.11.}
\label{fig-3a-r2}
\end{figure}

Another resonance, which was not calculated in \cite{fed96}, is clearly
seen in the spectrum with the energy $E=4.30-0.32i$~MeV, that is with
the excitation energy $E^*$=11~MeV and the width $\Gamma$=0.6~MeV. It
apparently corresponds to the experimentally known broader second excited
0$^+$ state with $E^*$=10.3$\pm$0.3~MeV and $\Gamma$=3$\pm$0.7~MeV
\cite{aizenberg}.

The dominating radial function, which for this resonance corresponds
to the first eigenvalue on Fig.~\ref{fig-3a-l}, is shown on
Fig.~\ref{fig-3a-r2}.  The real part of the wavefunction has no nodes
in the pocket region.  Therefore this resonance represents the lowest
state in the hyper-angular eigenvalue number 1. This eigenvalue also
forms a pocket, see Fig.~\ref{fig-3a-l}, although not as deep as that
of eigenvalue number 2.

\section{Conclusion}

We have implemented complex scaling of coordinates within the
hyper-spheric adiabatic approach to Faddeev equations.  We have derived
the necessary equations and investigated the asymptotic spectrum of the
hyper-angular eigenvalues.

The complex-scaled hyper-radial wave-functions fall off exponentially at
large distances which allows the simple boundary condition of vanishing
functions to be used in the search for the resonances.
Alternatively, it is possible to use square integrable functions as a
basis set, thus avoiding the less effective shoot-and-match method. Using
such a basis expansion has an additional advantage that all bound states
and resonances are calculated in one diagonalization run while otherwise
one has to search separately for every single pole of the S-matrix.
Again with basis expansion in addition to resonances and bound states
one also gets in the same run the (discretized) continuum wave functions.
Otherwise continuum wave functions have to be calculated separately.

The simplified boundary conditions are especially important for the three
positively charged particles where otherwise the non-scaled boundary
condition involves complicated functions with slow asymptotic convergence.

The drawbacks of the complex scaling method is first that it requires
the numerically slower complex arithmetics to be used and second that the
ranges of short-range potentials effectively become larger after complex
scaling with the numerically stronger demand of more accurate treatment
at larger distances.

We have illustrated the viability of the method by applications to
three systems, one benchmark model system and two realistic systems,
one with short-range potentials and the other in addition also including
the long-range Coulomb potentials.  In conclusion the complex-scaled
hyper-spheric approach to Faddeev equations is a viable alternative method
for calculating three-body resonances and continuum spectrum states.

\end{document}